\documentclass[aps,prb,twocolumn,superscriptaddress,showpacs,amsmath,showpacs]{revtex4}
\usepackage{graphicx,amssymb}

\begin{document}

%

\title{Quantum dynamics of the N\'eel vector in the antiferromagnetic molecular wheel CsFe$_8$}

\author{O. Waldmann}
 \email[Corresponding author.\\E-mail: ]{waldmann@iac.unibe.ch}
 \affiliation{Department of Chemistry and Biochemistry, University of Bern, CH-3012 Bern, Switzerland}

\author{C. Dobe}
 \affiliation{Department of Chemistry and Biochemistry, University of Bern, CH-3012 Bern, Switzerland}

\author{H. U. G\"udel}
 \affiliation{Department of Chemistry and Biochemistry, University of Bern, CH-3012 Bern, Switzerland}

\author{H. Mutka}
 \affiliation{Institut Laue-Langevin, 6 rue Jules Horowitz, BP 156, 38042 Grenoble Cedex 9, France}

\date{\today}

\begin{abstract}
The inelastic neutron scattering (INS) spectrum is studied for the antiferromagnetic molecular wheel
CsFe$_8$, in the temperature range 2 - 60~K, and for transfer energies up 3.6~meV. A qualitative analysis
shows that the observed peaks correspond to the transitions between the $L$-band states, from the ground
state up to the $S$ = 5 multiplet. For a quantitative analysis, the wheel is described by a microscopic spin
Hamiltonian (SH), which includes the nearest-neighbor Heisenberg exchange interactions and uniaxial easy-axis
single-ion anisotropy, characterized by the constants $J$ and $D$, respectively. For a best-fit determination
of $J$ and $D$, the $L$ band is modeled by an effective SH, and the effective SH concept extended such as to
facilitate an accurate calculation of INS scattering intensities, overcoming difficulties with the dimension
of the Hilbert space. The low-energy magnetism in CsFe$_8$ is excellently described by the generic SH used.
The two lowest states are characterized by a tunneling of the N\'eel vector, as found previously, while the
higher-lying states are well described as rotational modes of the N\'eel vector.
\end{abstract}


\maketitle

%

\section{Introduction}

Molecular nanomagnets have recently attracted enormous attention, as these perfect magnetic nanoclusters were
demonstrated to exhibit a number of spectacular quantum phenomena. In systems such as the single-molecule
magnets Mn$_{12}$ and Fe$_8$, quantum tunneling of the magnetization, and related effects such as quantum
interference, were observed.\cite{Ses93,Fri96,Tho96,Wer99,Mn12_Fe8} These phenomena are intimately connected
to a high-spin ground state, so that at low temperatures these clusters essentially behave like large single
spins $S$, where $S$ = 10 for Mn$_{12}$ and Fe$_8$. This single-spin (or giant-spin) model is in fact
extremely successful in describing the magnetism of single-molecule magnets.\cite{Mn12_Fe8} The actual
underlying many-spin nature of the system, despite providing a large $S$, is not key to the observed
tunneling phenomena.

\begin{figure}[b]
\includegraphics{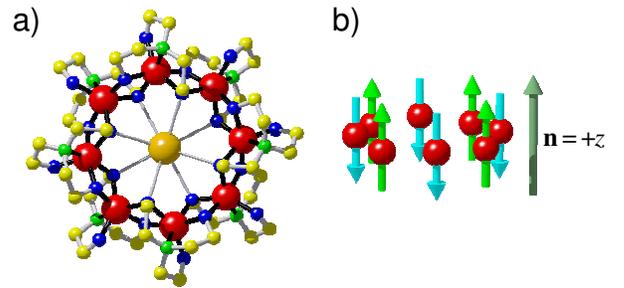}
\caption{\label{fig1} (Color online) (a) Crystal structure of CsFe$_8$ [with Fe as dark gray spheres (red
online) and H atoms omitted]. (b) Classical ground-state spin configuration for the N\'eel vector [long up
arrow (dark green online]) oriented in the $+z$ direction.}
\end{figure}

Another class of interesting molecules are the antiferromagnetic (AFM) wheels, in which the magnetic centers
are arranged into almost perfect rings (only wheels with an even number of equivalent spin centers shall be
considered).\cite{Taf94,Gat94,Inn05,OW_CCR} A variety of quantum phenomena were observed, such as
magnetization steps,\cite{Taf94,Cor99,Cor99b,OW_XFe6,OW_CsFe8,Aff02} quantized rotation of the N\'eel vector
or quantized spin waves,\cite{OW_Cr8} and quantum tunneling of the N\'eel vector.\cite{OW_NVT_CsFe8,San05} In
AFM wheels, in contrary to single-molecule magnets, the ground state has a total spin of $S$ = 0 and hence is
nonmagnetic. A single-spin model (with $S$ = 0) would thus be inappropriate as it would not allow one to
describe the interesting spin dynamics in these systems, which are manifestly many-spin effects. The
lowest-lying excitations can be described in terms of the N\'eel vector, $\textbf{n}$, which is essentially a
vector parallel to the magnetization of one of the AFM sublattices [in a classical picture it could be for
example parallel to the up-pointing spins as in Fig.~\ref{fig1}(b)].

The magnetism of the AFM wheels is very well described by the generic spin Hamiltonian
\begin{equation}
\label{H}
 \hat{H} = -J \left( \sum^{N-1}_{i=1}{ \hat{\textbf{S}}_i \cdot \hat{\textbf{S}}_{i+1} } +
\hat{\textbf{S}}_N \cdot \hat{\textbf{S}}_1 \right) + D \sum^N_{i=1} \hat{S}^2_{i,z},
\end{equation}
which includes isotropic (Heisenberg) nearest-neighbor exchange interactions, characterized by the coupling
constant $J < 0$, and an uniaxial magnetic anisotropy, of the easy-axis type, described by an on-site
anisotropy with $D < 0$.\cite{dipdip} $N$ is the number of spin centers, $\hat{\textbf{S}}_i$ is the spin
operator of the $i$th ion with spin $s$, and $z$ is the uniaxial anisotropy axis perpendicular to the plane
of the wheel.

At small magnetic anisotropy, the lowest-lying excitations are characterized by a (quantized) rotation of the
N\'eel vector, which in the quantum mechanical energy spectrum shows up as a low-lying rotational mode, the
$L$ band. This band is comprised of the lowest state in each $S$ sector, i.e., the lowest $S$ = 0, 1, 2,
\ldots states, and their energies increase quadratically with $S$ according to $E(S) \propto
S(S+1)$.\cite{And52,Ber92,Sch00,OW_SPINDYN,OW_CCR} At higher energies, further rotational modes exist,
collectively denoted as $E$ band, which correspond to the quantized spin-wave excitations expected in AFM
spin systems.\cite{And52,OW_SPINDYN,OW_CCR} This picture of the excitations has been demonstrated in great
detail experimentally by inelastic neutron scattering (INS) on the AFM wheel Cr$_8$ (for which $J$ =
-1.46~meV, $D$ = -0.038~meV, or $S_0/\hbar \equiv Ns\sqrt{2D/J} = 2.7$).\cite{OW_Cr8,DJ}

If the magnetic anisotropy becomes strong, and exceeds the threshold $S_0/\hbar > 4$,\cite{DJ} the situation
changes completely for the lowest lying excitation, which is then characterized by quantum tunneling of the
N\'eel vector.\cite{Bar90,Chi98,Mei01} This scenario has recently been observed by INS in the title compound,
the AFM wheel CsFe$_8$ [the structure is shown in Fig.~\ref{fig1}(a), here $J$ = -1.80~meV, $D$ = -0.05~meV,
or $S_0/\hbar = 4.7$].\cite{OW_NVT_CsFe8}

In this work, the AFM wheel CsFe$_8$ is further investigated by INS measurements, extending the previous
work, Ref.~\onlinecite{OW_NVT_CsFe8}, to a larger temperature and energy transfer range, which allows us to
probe a much larger part of the low-lying energy spectrum. In particular, the $L$-band states have been
detected from the ground state up to the $S$ = 5 multiplet at an energy of ca. 15~meV. This provides a
detailed insight into the dynamics of the N\'eel vector in the CsFe$_8$ wheel.

In the next section~II, the experimental details and the data are presented. The analysis of the data has
been split into two parts. Initially, a qualitative analysis is provided in section~III. Subsequently, in
section~IV some preparatory work for a quantitative analysis is developed, with the final analysis presented
in section~V. Section~VI provides a discussion of the results, and the manuscript concludes in section~VII.

%

\section{Experimental}

The CsFe$_8$ material, which has the chemical formula [CsFe$_8$L$_8$]Cl $\cdot$ 2CHCl$_3$ $\cdot$
0.5CH$_2$Cl$_2$ $\cdot$ 0.75L $\cdot$ 2.5H$_2$O, with L = N(CH$_2$CH$_2$O)$_3$, was prepared by the same
procedure as in Ref.~\onlinecite{Saa97}, but crystallized from a 1:1 mixture of CHCl$_3$ and CH$_2$Cl$_2$ by
pentane vapor diffusion. The compound crystallizes in the monoclinic space group Pna21, and the molecules
exhibit approximate C$_4$ symmetry with Fe-Fe distances of 3.142 - 3.164~\AA, Fig.~\ref{fig1}(a). The INS
spectrum of CsFe$_8$ was measured with the time-of-flight spectrometer IN5 at the Institute Laue-Langevin
(ILL), Grenoble, France. 4~g of a non-deuterated powder sample was sealed under a helium atmosphere in a
double-walled hollow Al cylinder (50~mm in height, 16~mm external diameter, and 2~mm thickness) and inserted
in a standard ILL orange $^4$He cryostat. INS spectra were recorded at temperatures between 2.2~K and 60~K
for incident neutron wavelengths of $\lambda$ = 3.8~{\AA} and 5.0~{\AA} (energy resolution at the elastic
peak was 161~$\mu$eV and 121~$\mu$eV, respectively). The data were corrected for detector efficiency using a
vanadium standard; no further corrections were applied. The data shown correspond to the sum over all
detector banks; the error bars are smaller than the symbol size.

\begin{figure}
\includegraphics{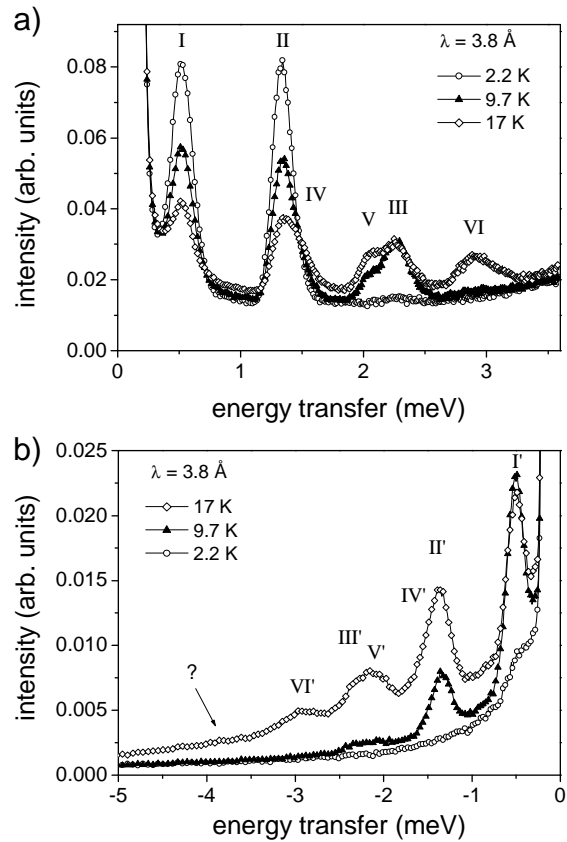}
\caption{\label{fig2} Inelastic neutron scattering spectra of CsFe$_8$ at $T$ = 2.2~K, 9.7~K, and 17~K,
measured with $\lambda$ = 3.8~{\AA}. Panel (a) shows the neutron energy-loss, and panel (b) the neutron
energy-gain side of the spectrum. The question mark indicates a weak feature which will be identified as a
further transition VII' later on.}
\end{figure}

\begin{figure}
\includegraphics{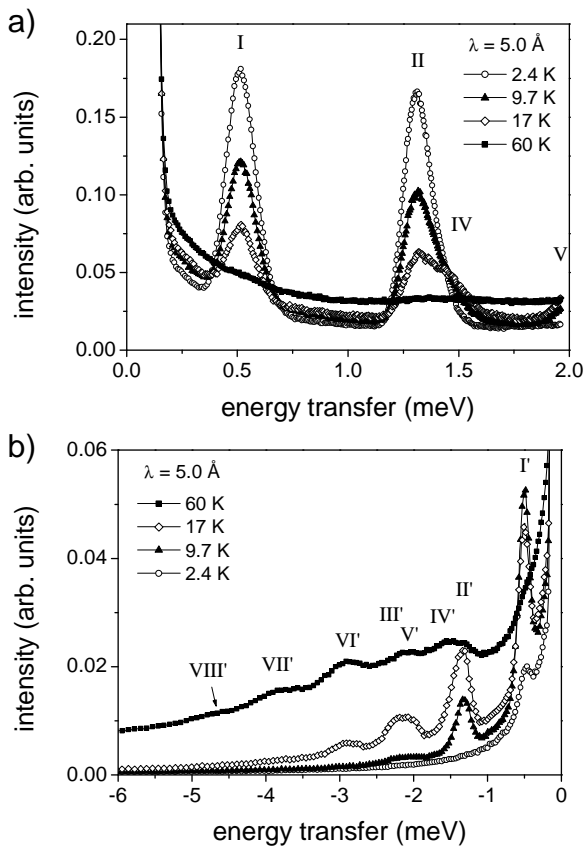}
\caption{\label{fig3} Inelastic neutron scattering spectra of CsFe$_8$ at $T$ = 2.4~K, 9.7~K, 17~K, and 60~K,
measured with $\lambda$ = 5.0~{\AA}. Panel (a) shows the neutron energy-loss, and panel (b) the neutron
energy-gain side of the spectrum.}
\end{figure}

Figure~\ref{fig2} presents the neutron energy-loss and energy-gain sides of the INS spectrum as measured at
$\lambda$ = 3.8~{\AA} for the temperatures $T$ = 2.2~K, 9.7~K, and 17~K. Figure~\ref{fig3} presents the
$\lambda$ = 5.0~{\AA} data obtained for the temperatures $T$ = 2.4~K, 9.7~K, 17~K, and 60~K.

%

\section{Qualitative analysis}

\begin{figure}
\includegraphics{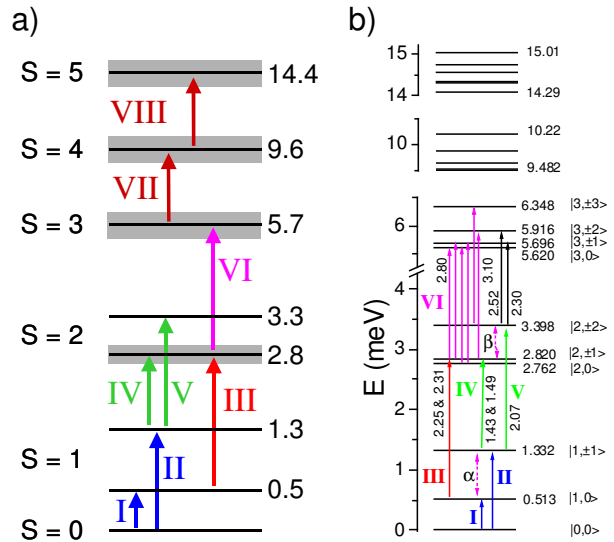}
\caption{\label{fig4}(Color online) Energy spectrum of CsFe$_8$ as (a) derived from the neutron scattering
data, and (b) calculated numerically from the microscopic Hamiltonian $\hat{H}$, eq. (\ref{H}), using the
best-fit parameters determined in this work. The level energies are given in meV. The labeling of the
transitions, the classification of the levels, and the transitions $\alpha$ and $\beta$ are discussed in the
text.}
\end{figure}

The interpretation of the INS spectra is relatively straightforward. In the 3.8~{\AA} neutron energy-loss
spectrum at 2.2~K, Fig.~\ref{fig2}(a), two peaks I and II at about 0.5~meV and 1.3~meV, respectively, are
observed. They correspond to transitions from the ground state to two next higher-lying levels at the
corresponding excitation energies, as indicated in Fig.~\ref{fig4}(a). At 9.7~K the intensities of peaks I
and II decrease, and a new transition III at 2.3~meV develops, which is assigned to a transition from the
first excited level to a higher-lying level at 2.8~meV. Some additional weaker features at 1.5~meV and
2.0~meV (peaks IV and V) are visible, which grow stronger at 17~K. They are assigned to transitions from the
second excited level to higher-lying levels at again 2.8~meV and 3.3~meV. At 17~K a further strong feature at
2.9~meV, peak VI, is observed. As it is essentially absent in the data at the lower temperatures, it is
assigned to a transition from the 2.8~meV level to a level at about 5.7~meV. On the neutron energy-gain side
of the data, Fig.~\ref{fig2}(b), all corresponding anti-Stokes peaks (labeled with an additional apostrophe)
are observed with the expected temperature dependence. This confirms the above assignment of the levels, also
demonstrating the magnetic origin of the transitions.

The 5.0~{\AA} neutron scattering data shown in Fig.~\ref{fig3} confirm the findings from the 3.8~{\AA}
measurements with higher resolution (the 2.4~K and 9.7~K data were communicated previously in
Ref.~\onlinecite{OW_NVT_CsFe8}). The 60~K spectrum provides valuable additional information. On the neutron
energy-loss side, Fig.~\ref{fig3}(a), resolved transitions are no longer observed; however, an increased
inhomogeneous intensity as compared to the 2.4~K data is evident. Apparently, at 60~K, a large number of
transitions with widely varying transition energies become possible, and individual transitions cannot be
resolved. We interpret this as the onset of transitions into the so-called quasi-continuum in the spectrum of
AFM molecular wheels.\cite{OW_SPINDYN,OW_CCR} In the neutron energy-gain spectrum at 60~K,
Fig.~\ref{fig3}(b), the anti-Stokes peaks I' to VI' are clearly visible, as well as two further transitions
VII' and VIII' at 3.9~meV and 4.8~meV, respectively (the presence of peak VIII' will be unambiguously
confirmed in Fig.~\ref{fig6} below). Very weak indications of peak VII' can be found in both the 3.8~{\AA}
and 5.0~{\AA} neutron energy-gain data at 17~K. This peak is thus assigned to a transition from the excited
5.7~meV level to a further higher-lying level at about 9.6~meV. Based on the weaker intensity of peak VIII',
it is assigned to a transition from the 9.6~meV level to a level at about 14.4~meV. The magnetic origin of
peaks VII' and VIII', and their assignment, is less evident from the data, but is confirmed by the
quantitative analysis presented in section~IV.

\begin{figure}
\includegraphics{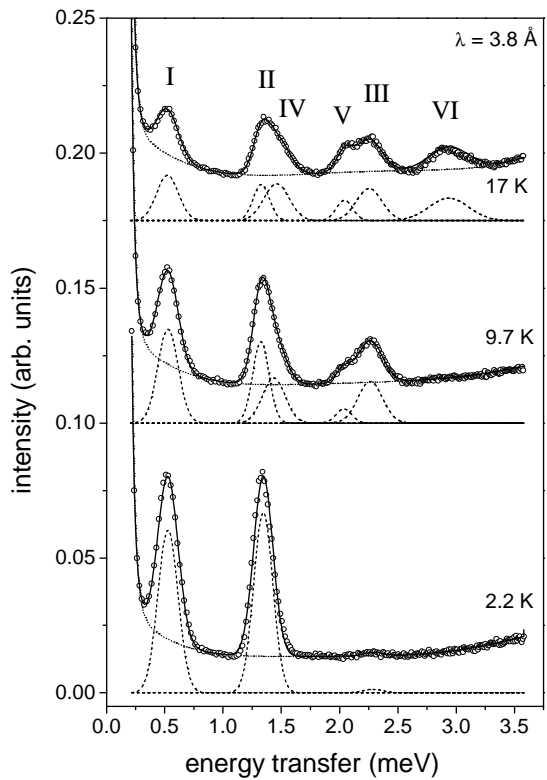}
\caption{\label{fig5} Analysis of the $\lambda$ = 3.8~{\AA} neutron energy-loss spectra for $T$ = 2.2~K,
9.7~K, and 17~K. The data curves for $T$ = 9.7~K and 17~K were shifted for clarity by 0.1 and 0.175
arb.~units, respectively. Open circles represent the data and solid lines fits, which included an
approximated background (dotted lines) and a number of Gaussian lines (dashed lines). The parameters of the
Gaussian lines are given in Table~A.II.}
\end{figure}

With the above thermal-population arguments the energy spectrum as compiled in Fig.~\ref{fig4}(a) is directly
concluded from the INS measurements. For the further analysis, the following procedure was applied to each
data curve on the neutron energy-loss side. The background was approximated by an analytical function and
substracted from the data. The corrected spectrum was then least-squares fitted to an appropriate number of
Gaussian peaks. The decomposition into the different contributions is explicitly shown for the 3.8~{\AA} data
in Fig.~\ref{fig5}; the fitting results for the peak positions, linewidths, and intensities are compiled in
the Tables given in appendix~A. The temperature variation of the peak intensities are in accord with the
expected thermal Boltzmann population, supporting the above analysis.

It is interesting to inspect the linewidths provided in Tables~A.I and A.II. The linewidths of transitions I,
II, and V are consistent with the experimental resolution for both the 3.8~{\AA} and 5.0~{\AA} data sets.
This indicates that the energy levels at 0.5~meV, 1.3~meV, and 3.3~meV correspond to single states in the
energy spectrum of CsFe$_8$. The linewidths of transitions III and IV, which are about 0.21~meV (FWHM) for
both peaks, are consistently larger than the experimental resolution. This indicates that the level at
2.8~meV in Fig.~\ref{fig4}(a) actually relates to two (or more) close-lying states in the energy spectrum.
Finally, the linewidth of transition VI, ca. 0.34~meV (FWHM), is even larger, which indicates that also the
level at 5.7~meV contains two or more close-lying energy states. These findings are represented in
Fig.~\ref{fig4}(a) by the thicker gray bars.

\begin{figure}
\includegraphics{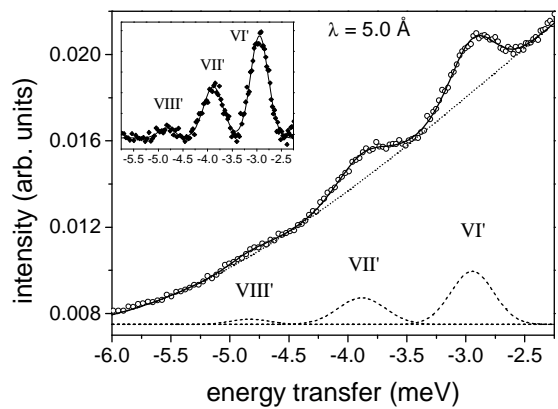}
\caption{\label{fig6} Analysis of a part of the $\lambda$ = 5.0~{\AA} neutron energy-gain spectra at $T$ =
60~K. Open circles represent the data and the solid line the best fit, which included an approximated
background (dotted lines) and three Gaussian lines (dashed lines). The parameters of the Gaussian lines are
given in Table~A.III. The inset shows the data after substraction of the background (black squares) and the
three fitted Gaussian lines (solid line).}
\end{figure}

A similar analysis was performed for the 5.0~{\AA} neutron energy-gain data at 60~K in an energy range which
embraces peaks VI', VII', and VIII'. The decomposition of the data is shown in Fig.~\ref{fig6}; the best-fit
values for the three Gaussian lines are given in Table~A.III. The inset of Fig.~\ref{fig6} confirms the
statistical significance of peak VIII'. The observed linewidths of 0.33-0.4~meV are significantly larger than
the experimental resolution, with similar conclusions as for the 3.8~{\AA} data.

A qualitative assignment of the observed transitions is achieved by comparing the level scheme in
Fig.~\ref{fig4}(a) with the energy spectrum as calculated numerically from $\hat{H}$, eq.~(\ref{H}). In order
to calculate the energy spectrum, one could use the $J$ and $D$ values inferred previously from high-field
torque measurements.\cite{OW_CsFe8} Since these values, however, are very close to the best-fit values to be
determined below in section~V, the energy spectrum obtained for the best-fit values, which is presented in
Fig.~\ref{fig4}(b), will be used here. In Fig.~\ref{fig4}(b), the states are labeled by the quantum numbers
$S$ and $M$ of the total-spin operator $\hat{\textbf{S}} = \sum_i \hat{\textbf{S}}_i$ (using the notation
$|S,M\rangle$ for the levels). However, it should be stressed here that $S$ is not a good quantum number for
CsFe$_8$, because of the rather large magnetic anisotropy present in the system.\cite{OW_NVT_CsFe8} For the
moment, $S$ should be taken as an arbitrary additional label for distinguishing the energy levels; a further
detailed discussion of the situation will be given in section~VI.

From a comparison of Figs.~\ref{fig4}(a) and \ref{fig4}(b), the identification of the experimentally observed
transitions is apparent, considering the INS selection rules $\Delta S = 0,\pm1$ and $\Delta M = 0,\pm1$. The
0.5~meV and 1.3~meV levels correspond to transitions from the ground state $|0,0\rangle$ to the levels
$|1,0\rangle$ and $|1,\pm1\rangle$, respectively. As demonstrated in Ref.~\onlinecite{OW_NVT_CsFe8}, peak I
reflects the quantum tunneling of the N\'eel vector characterizing the low-energy spin dynamics in CsFe$_8$.
It was also noted that the transition between the first and second excited levels (denoted as $\alpha$) is
not observed, although it is formally allowed by the INS selection rules. The intensity of $\alpha$ is very
weak compared to that of the other transitions, since the sublattice-magnetization vectors are of mesoscopic
size in CsFe$_8$. Peaks III, IV, and V correspond to transitions from the $S=1$ states to those of the $S=2$
multiplet. Interestingly, peaks III and IV are confirmed to consist of two close transitions, in agreement
with the observed enlarged linewidths (the transition $|2,\pm1\rangle \rightarrow |2,\pm2\rangle$, which is
expected at ca. 0.6~meV at higher temperatures, turns out to be too weak in intensity to be detectable). Peak
VI corresponds to transitions from the $S=2$ states into the $S=3$ multiplet. The zero-field-splitting of
these multiplets, which allows six close transitions, is reflected by the broad linewidth of peak VI.
Finally, peaks VII and VIII correspond to transitions from the $S=3$ to the $S=4$, and the $S=4$ to the $S=5$
multiplets, respectively. Here again, the spread in the transition energies of the individual transitions due
to the zero-field-splitting is too small to be resolved experimentally, hence the broadened peaks.

The energy level diagram shown in Fig.~\ref{fig4}(a) is one of the most extended level schemes determined for
a molecular nanomagnet from INS measurements, yet all essential features of the data can be readily explained
by basic arguments. The above analysis, in our opinion, provides a nice textbook example for the analysis of
INS data from molecular clusters.

%

\section{Effective Hamiltonian and inelastic neutron scattering intensity}

In the next section a more sophisticated analysis of the data will be presented, which yields best-fit values
for $J$ and $D$, as well as some insight into the accuracy and limitations of the generic spin Hamiltonian
(\ref{H}). A fit of the data to INS spectra calculated from the microscopic Hamiltonian (\ref{H}) using
standard least-squares fitting routines, however, is numerically expensive for a system as large as CsFe$_8$.
Another procedure has thus been developed, the essentials of which are outlined in this section.

In our experiments, only transitions between states of the so-called $L$ band were
observed.\cite{OW_SPINDYN,OW_Cr8,OW_CCR} The $L$ band, also known as the tower of states\cite{And52} or the
quasi-degenerate joint states,\cite{Ber92,Lec97} arises from a quantized rotation of the N\'eel vector in AFM
wheels and other systems. It consists of the energetically lowest-lying states for each $S$, whose energies
increase as $E(S) \propto S(S+1)$. This set of states can be very well described for small AFM Heisenberg
rings by approximating the wave functions by $|\beta_A \beta_B S_A S_B S M\rangle$, with $S_A=S_B=10$ for
CsFe$_8$ ($S_A$ and $S_B$ denote the total spin of sublattices $A$ and $B$, respectively. $\beta_A$ and
$\beta_B$ abbreviate intermediate quantum numbers, but are omitted in the
following).\cite{OW_SPINDYN,OW_FW_QT} Physically, this approach works well because the internal spin
structure due to the dominant Heisenberg interaction is essentially classical.\cite{OW_Cr8,OW_CCR} In this
approximation, the effective (two-sublattice) Hamiltonian
\begin{equation}
\label{Heff}
 \hat{H}_{AB} = - \tilde{J} \hat{\textbf{S}}_A \cdot \hat{\textbf{S}}_B + \tilde{D} ( \hat{S}^2_{A,z} + \hat{S}^2_{B,z} )
\end{equation}
with $\tilde{J} = 0.5366 J$ and $\tilde{D} = 0.1870 D$ is obtained for CsFe$_8$.\cite{OW_FW_QT} Since
$\hat{H}_{AB}$ is designed to describe only the states of the $L$ band, numerical calculations become much
cheaper (for CsFe$_8$, the dimension of the Hilbert space is 1679616 while that of $\hat{H}_{AB}$ is only
441).

Concerning the energies, $\hat{H}_{AB}$ has been demonstrated to yield very accurate results.\cite{OW_FW_QT}
The calculation of INS spectra, however, needs further consideration since not only the energies but also the
local transition matrix elements and so-called interference factors need to be properly approximated. While
transition matrix elements are expected to be reproduced reasonably well, the situation is less obvious for
the interference factors. In fact, these factors, which typically produce a pronounced oscillatory behavior
of the INS scattering intensity as function of momentum transfer ($Q$ dependence), arise from a correlation
of the transition matrix elements and the geometrical structure of the molecule.\cite{Fur77,OW_INS}
$\hat{H}_{AB}$ essentially replaces the octanuclear spin-5/2 ring structure of CsFe$_8$ by a dimer of two
spin-10 centers, which is obviously a very different geometrical structure. A naive calculation of the INS
spectra from $\hat{H}_{AB}$ should thus be suspected to produce incorrect $Q$ dependencies, and hence
incorrect INS intensities. For CsFe$_8$ this is indeed the case. In the following, however, it will be shown
that the interference effects can be retained by a suitable generalization of the INS scattering formula. The
procedure is actually very simple.

For a spin cluster, the INS cross section is given by\cite{Fur79,Gue85}
\begin{equation}
\label{cross_section}
 {d^2 \sigma \over d\Omega d\omega} = C(Q,T)
\sum_{nm} {e^{-\beta E_n} \over Z(T)} I_{nm}({\bf Q}) \delta(\omega-{E_m-E_n\over \hbar}),
\end{equation}
where $C(Q,T)=(\gamma e^2/m_e c^2) (k'/k) \exp[-2 W(Q,T)]$ (where all symbols have the usual meaning), $\beta
= 1/(k_B T)$, and $Z(T)$ is the partition function. For a powder sample in zero magnetic field one
finds\cite{OW_INS}
%
\begin{eqnarray}
\label{Inm}
 I_{nm}(Q) &=& \sum_{ij} F^*_i(Q) F_j(Q) \{ {2 \over 3} j_0(Q R_{ij}) {\bf \tilde{S}}_i \cdot {\bf \tilde{S}}_j
 \cr &&
  + j_2(Q R_{ij}) \sum_q T^{(2)*}_q({\bf R}_{ij}) T^{(2)}_q( {\bf \tilde{S}}_i{\bf \tilde{S}}_j) \},
\end{eqnarray}
%
where $F_i(Q)$ is the magnetic form factor of the $i$th spin center, $j_k$ is the spherical Bessel function
of order $k$, and ${\bf R}_{ij} = {\bf R}_{i} - {\bf R}_{j}$ is the distance vector between the $i$th and
$j$th ion. An explicit expression for Eq.~(\ref{Inm}) is given in Ref.~\onlinecite{OW_INS2}. $T^{(k)}_q({\bf
v})$ is the $q$th component of the spherical tensor of rank $k$ constructed from the Cartesian vector ${\bf
v}$, and $T^{(2)}_q( {\bf \tilde{S}}_i{\bf \tilde{S}}_j)$ represents the tensor product $[T^{(1)}({\bf
\tilde{S}}_i) \otimes T^{(1)}({\bf \tilde{S}}_j)]^{(2)}_q$. The ordered products $\tilde{S}_{i \alpha}
\tilde{S}_{j \beta}$, which appear in the explicit expression of $T^{(2)}_q( {\bf \tilde{S}}_i{\bf
\tilde{S}}_j)$, stand for $\langle n|\hat{S}_{i \alpha}|m\rangle\langle m| \hat{S}_{j \beta}|n\rangle$, where
$|n\rangle$ denotes an eigenstate of the Hamiltonian under consideration. Equation~(\ref{Inm}) can be also
written more compactly as
\begin{eqnarray}
\label{Inm2}
 I_{nm}(Q) = \sum_{ij} \sum_{kq} f^{kq}_{ij}(Q,{\bf R}_{ij}) U^{(k)}_q({\bf
\tilde{S}}_i {\bf \tilde{S}}_j),
\end{eqnarray}
whereby introducing the interference factors $f^{kq}_{ij}(Q,{\bf R}_{ij})$ ($k = 0, 2$ and $|q| \leq k$) and
the symmetrized spherical tensors $U^{(k)}_q$ (which are proportional to Re[$T^{(k)}_q$] for $q \geq 0$ and
Im[$T^{(k)}_q$] for $q < 0$, see appendix~B).

The easiest way to derive an effective formula for the INS intensity is to follow the procedure suggested in
Ref.~\onlinecite{OW_NVT_3x3} for the construction of an effective Hamiltonian for AFM spin clusters with
essentially classical internal spin structure. It starts by noting that a classical spin structure relates to
a mean-field situation for each of the AFM sublattices. Accordingly, all spins $\hat{{\bf S}}_i$ on a
sublattice $\nu$ are replaced by the mean-field spin $\hat{{\bf S}}_{\nu}/N_{\nu}$, where $\hat{{\bf
S}}_{\nu} = \sum_{i \in \nu} \hat{{\bf S}}_i$ and $N_{\nu}$ is the number of spins on sublattice $\nu$.

Concerning the INS intensity, one replaces all matrix elements $\tilde{S}_{i \alpha}$ with $i \in \nu$ by
$\tilde{S}_{\nu \alpha}/N_\nu$, puts all $\tilde{S}_{\nu \alpha}$ outside the brackets, and immediately
obtains
\begin{eqnarray}
\label{Ieff}
 I_{nm}(Q) = \sum_{\nu\mu} \sum_{kq} \bar{f}^{kq}_{\nu\mu}(Q) U^{(k)}_q({\bf \tilde{S}}_\nu {\bf \tilde{S}}_\mu)
\end{eqnarray}
with the effective interference factors
\begin{eqnarray}
\label{F}
 \bar{f}^{kq}_{\nu\mu}(Q) = {1 \over N_\nu N_\mu} \sum_{i \in \nu, j \in \mu} f^{kq}_{ij}(Q,{\bf R}_{ij}).
\end{eqnarray}

In the case of CsFe$_8$, with its two sublattices $A$ and $B$, Eq.~(\ref{Ieff}) yields a scattering formula
analogous to that of a dimer, but with the dimer interference factors $f_{11}(Q)$ and $f_{12}(Q)$ replaced by
the effective interference factors $\bar{f}_{AA}(Q)$ and $\bar{f}_{AB}(Q)$. In the scattering formula, now
only the matrix elements $\tilde{S}_{A \alpha}$ and $\tilde{S}_{B \alpha}$ appear, which can be calculated
from the eigenpairs obtained from $\hat{H}_{AB}$.

\begin{figure}
\includegraphics{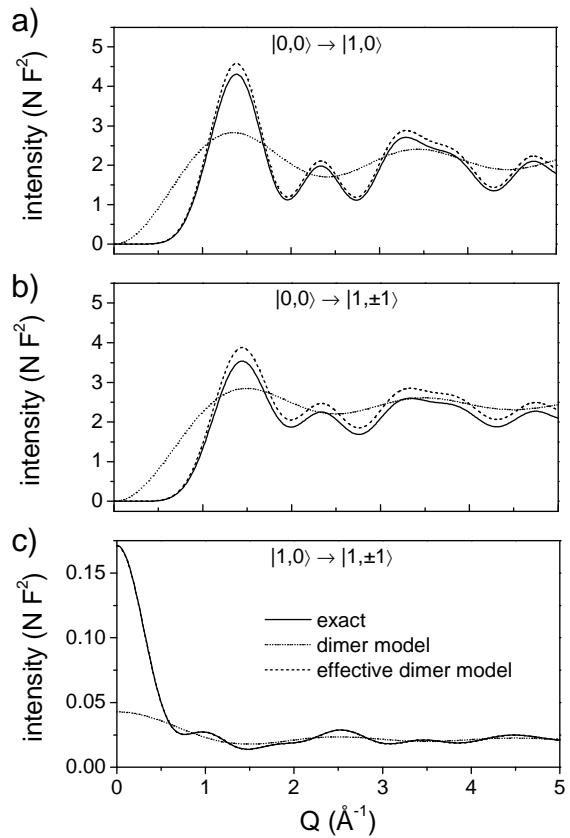}
\caption{\label{fig7} $Q$ dependence of the neutron scattering intensity of CsFe$_8$ as calculated from the
full Hamiltonian (\ref{H}) (solid lines), the dimer Hamiltonian $\hat{H}_{AB}$ (dotted lines), and the dimer
Hamiltonian $\hat{H}_{AB}$ using the effective INS formula Eq.~(\ref{Ieff}) (dashed lines). Results are shown
for the three transitions (a) $|0,0\rangle \rightarrow |1,0\rangle$, (b) $|0,0\rangle \rightarrow
|1,\pm1\rangle$, and (c) $|1,0\rangle \rightarrow |1,\pm1\rangle$, see Fig.~\ref{fig4}. In the calculations
with the pure dimer model, the distance between the two spins $S_A$ and $S_B$ was set to 3.2~{\AA},
corresponding to the iron-iron distance in CsFe$_8$.}
\end{figure}

Figure~\ref{fig7} presents numerical results for the $Q$ dependence of the INS intensity in CsFe$_8$ as
calculated from the microscopic spin Hamiltonian $\hat{H}$ (solid lines), the effective spin Hamiltonian
$\hat{H}_{AB}$ (dotted lines), and the effective Hamiltonian $\hat{H}_{AB}$ but now using Eq.~(\ref{Ieff})
for the INS intensity (dashed lines) [in the first two cases Eq.~(\ref{Inm}) was used]. Noting
Fig.~\ref{fig7} it is clear that the pure dimer model is not capable of producing correct $Q$ dependencies.
This is not surprising, and just reflects the fact that a dimer model can only describe the interference
between two ions, while in the full cluster interference occurs between eight ions. This also explains why
the exact $Q$ dependence shows more oscillations per $Q$ range. In contrast, the above procedure of
reinstating the full geometrical structure of the molecule works very well. For the intra-multiplet
transition $\alpha$ ($|1,0\rangle \rightarrow |1,\pm1\rangle$), the exact and effective dimer curves are
essentially indistinguishable in Fig.~\ref{fig7}(c). The curves in Figs.~\ref{fig7}(a) and (b) also coincide
with each other if one scales the effective dimer curves by factors of 0.940 and 0.911, respectively. These
numbers also give an idea of the accuracy of the transition matrix elements obtained from the effective
Hamiltonian approach, which apparently is better than 10\%. Furthermore, Fig.~\ref{fig7} clearly shows that
for $Q > 0.5$~{\AA} the intensity of the intra-spin-multiplet transition $\alpha$ is indeed two orders of
magnitude weaker than that of transitions I and II as noted previously.\cite{OW_NVT_CsFe8}

%

\section{Quantitative analysis}

In this section a complete quantitative analysis of the experimental data is provided. Typically,
spin-Hamiltonian parameters are obtained by least-square fitting the calculated energies (or transition
energies) to the peak positions obtained from a Gaussian analysis of the data, similar to that shown in
Fig.~\ref{fig5}. However, since in the present case some of the observed peaks are composed of several
individual but close-lying transitions the experimental spectra were fitted directly by simulated INS
spectra.

Only the results of the analysis for the 3.8~{\AA} neutron energy-loss data at 17~K will be shown explicitly,
as this data set contains the most information and thus allows the best judgment as to the agreement between
experiment and theory. A similar analysis for the other data yields redundant results. For each pair of
values for $J$ and $D$, theoretical INS spectra were calculated from the eigenpairs of $\hat{H}_{AB}$ using
the effective INS formula Eq.~(\ref{Inm2}). For the final best-fit values, and some other cases, spectra were
calculated from the microscopic spin Hamiltonian $\hat{H}$, using Eq.~(\ref{Inm}), in order to confirm the
accuracy of the results. In the calculations, the crystallographic positions of the iron(III) ions were used,
and the form factors $F_i(Q)$ were estimated by the standard analytical approximations.\cite{FFILL} The
variation of the $Q$ range with energy transfer $\hbar\omega$ of the IN5 spectrometer was accounted for by
numerically integrating $I(\omega,Q)$, where $Q \equiv Q(\omega,\theta)$, over the scattering angle spanned
by the detectors. $Q(\omega,\theta)$ is related to the energy transfer $\hbar\omega$ and the angle $\theta$
of the scattered neutron beam via the scattering triangle.

\begin{figure}
\includegraphics{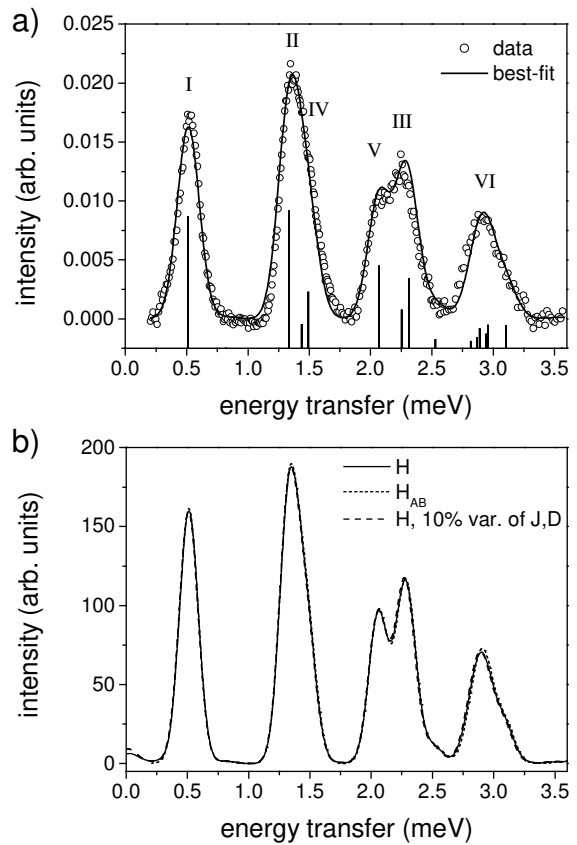}
\caption{\label{fig8} (a) Best-fit curve, to the background-substracted data, for the $\lambda$ = 3.8~{\AA}
neutron energy-loss spectrum for $T$ = 17~K ($J$ = -1.80~meV, $D$ = -0.050~meV). The vertical lines indicate
the transitions and their intensities which contribute to the neutron spectrum. (b) Calculated neutron
energy-loss spectra. The solid curve was calculated with the best-fit parameters using the effective
Hamiltonian approach, Eq.~(\ref{Ieff}), the short-dashed curve was calculated with the best-fit parameters
using the full Hamiltonian, Eq.~(\ref{Inm}), and the dashed curve was calculated from the full Hamiltonian
with a 10\% distribution in $J$ and $D$ (see text for details).}
\end{figure}

The result of a least-squares fit to the 3.8~{\AA}, 17~K neutron energy-loss data is shown in
Fig.~\ref{fig8}(a). The agreement is excellent, with the calculated curve reproducing all the details present
in the data. The obtained best-fit parameters were $J$ = -1.80(2)~meV and $D$ = -0.050(1)~meV, which are in
agreement with previous findings.\cite{OW_CsFe8,OW_NVT_CsFe8} The linewidth was set as a free parameter in
the fit, yielding 0.18~meV, very close to the nominal instrumental resolution of 0.161~meV. The good
agreement between experimental and calculated curves is also notable because several transitions contribute
to the various peaks, as indicated by the vertical solid lines in Fig.~\ref{fig8}(a). Peaks I, II, and V
correspond to individual transitions, while peaks III and IV are each made up of two close-lying transitions,
and peak VI is the result of six close-lying transitions, in perfect agreement with the conclusions drawn
from the analysis of the linewidths in section~III [see also Fig.~\ref{fig4}(a)]. The explicit assignment of
all the transitions which form the spectrum in Fig.~\ref{fig8} is given in Ref.~\onlinecite{peaks}. From the
expected transitions between the $S$ = 0, 1, 2, and 3 multiplets, all but two are observed. These are the two
intra-multiplet transitions $|2,\pm1\rangle\rightarrow|2,\pm2\rangle$ and
$|1,0\rangle\rightarrow|1,\pm1\rangle$ expected at 0.571~meV and 0.818~meV, respectively. Their intensity is
two orders of magnitude weaker as compared to the other transitions, because, as mentioned already and
discussed in Ref.~\onlinecite{OW_NVT_CsFe8}, the sublattice-magnetization vectors are of mesoscopic size in
CsFe$_8$. The other intra-multiplet transitions, while allowed, occur at too low energy and are thus buried
under the elastic line.

Using the best-fit parameters, the transitions between the $S$ = 3, 4, and 5 multiplets can also be
calculated. The result agrees nicely with the peak positions of peaks VII and VIII, thus confirming their
magnetic origin, and their assignment as transitions from the $S$ = 3 to $S$ = 4 and $S$ = 4 to $S$ = 5
multiplets, respectively.

As a summary of the analysis so far, the transitions between the $S$ = 0, 1, 2, 3, 4, and 5 multiplets (of
the $L$ band) were observed and could be reproduced very well by the Hamiltonian $\hat{H}$. Apparently, the
generic Hamiltonian $\hat{H}$, which contains only the two free parameters $J$ and $D$, provides an excellent
description of the excitations in CsFe$_8$, within the experimental resolution of the present experiments of
about 0.1~meV.

Indications have been found in previous works on other AFM wheels that Hamiltonian~(\ref{H}), though being an
excellent approximation, is not complete and needs to be extended by further weak
terms.\cite{Cor99b,Aff02,Cin02,OW_NaFe6,Car03,San05} This raises the question as to how well
Hamiltonian~(\ref{H}) in fact describes CsFe$_8$. It should be noted that for some wheels, effects which are
due to coupling of the spin system to the environment have been found,\cite{OW_NaFe6,OW_FIMEI} however these
are a different subject and not of interest here. The presence of further weak terms seems particularly
relevant for CsFe$_8$ as it does not exhibit eight-fold symmetry; the molecule investigated in this work has
a crystallographic $C_2$ axis. Accordingly, two deviations from the perfect situation described by
Hamiltonian~(\ref{H}) were considered, i) an additional biaxial anisotropy term and ii) a variation of the
exchange and anisotropy constants along the ring.

Biaxial anisotropy contributions for rather symmetric wheels were recently deduced for the Cr$_8$ and the
Fe$_{10}$ wheels.\cite{Car03,San05} In these cases, the microscopic Hamiltonian $\hat{H}_1 = \hat{H} + E
\sum_i (\hat{S}_{i,x}^2 - \hat{S}_{i,y}^2)$ was used. The Cr$_8$ wheel exhibits a crystallographic $C_4$
symmetry (though indications of a structural phase transition at low temperatures seem to have been found),
however the INS data were significantly better described with $|E|$ = 0.004(1)~meV, corresponding to a ratio
$|E/D|$ = 0.11.\cite{Car03} For the Fe$_{10}$ wheel, which exhibits crystallographic $C_2$ symmetry, $|E|$ =
0.0036(2)~meV was found, corresponding to a quite substantial ratio $|E/D|$ = 0.21.\cite{San05}

In the case of CsFe$_8$, an upper limit for the value of $|E|$ can be inferred from the present data as
follows. A biaxial anisotropy would split the two $|1,\pm 1\rangle$ levels, which are degenerate in a
uniaxial model. Accordingly peak II should split into two [see also Fig.~\ref{fig4}(a)]. The 5.0~{\AA} data
at low temperatures yields a linewidth for peak II of ca. 0.13~meV (see Table~A.II), very close to the
nominal experimental resolution. Thus, a splitting of, for example, 0.08~meV (corresponding to an estimated
linewidth of peak II of ca. 0.18~meV) would have been clearly detected in this experiment. Since the
splitting of the $|1,0\rangle$ and $|1,\pm 1\rangle$ levels by the $D$ term is ca. 0.8~meV, one thus can
conclude that $|E/D| < 0.1$, or with the determined value for $D$ that $|E| <$ 0.005~meV. For a more
quantitative analysis, the 3.8~{\AA}, 17~K neutron energy-loss data was least-squares fitted to the
effective-Hamiltonian version of $\hat{H}_1$, with $J$, $D$, and $E$ (and the linewidth) as free parameters.
Independent of the starting values, the fit converged to the same values for $J$ and $D$ as above, and to
$|E| < 5\times10^{-5}$~meV, that is, values which are zero within the statistical error. The inclusion of a
biaxial term thus did not improve the quality of the fit.

A more suitable technique for detecting a possible $E$ term, or other additional terms in the Hamiltonian, is
electron paramagnetic resonance (EPR), as it allows one to detect the transitions within a spin multiplet
with higher resolution (in the case of CsFe$_8$ this is ca. 0.01~meV compared to ca. 0.1~meV for the INS
experiment). EPR experiments at 35~GHz and 190~GHz on single-crystals of CsFe$_8$ were thus performed (and
will be reported elsewhere\cite{EPR}). The EPR spectra are well described using only uniaxial anisotropy,
which sets a rather stringent upper limit on the strength of a biaxial anisotropy. Thus from a magnetic
perspective CsFe$_8$ is a very uniaxial and symmetric wheel.

As a second possible deviation from the generic situation, a variation of the exchange and anisotropy
constant within the ring was considered, as described by the Hamiltonian $\hat{H}_2 =  - \sum_i J_i
\hat{\textbf{S}}_i \cdot \hat{\textbf{S}}_{i+1} + \sum_i D_i \hat{S}^2_{i,z}$ (there enforcing $\sum_i J_i =
J$ and $\sum_i D_i = D$). A variation might originate, for instance, from the reduced $C_2$ symmetry of the
ring, which gives rise to four slightly different Fe-Fe distances, or from disorder in the crystallites. In
order to test such an effect, $J_i$ and $D_i$ values were drawn from a Gaussian distribution with a 10\%
standard deviation, and the INS spectra calculated. This is shown in Fig.~\ref{fig8}(b), and is compared to
the calculation for the generic Hamiltonian~(\ref{H}) [Fig.~\ref{fig8}(b) also presents the INS spectrum as
calculated with the effective Hamiltonian approach, just to reinforce the high accuracy of this procedure].
It is apparent that a 10\% variation of the parameters is quite possible within the present experiments. This
is not unexpected because, as discussed in Ref.~\onlinecite{OW_CsFe8}, a change of the coupling and
anisotropy parameters along the ring has no effect on the energies of the states of the $L$ band up to first
order. However, this is not so for the states of the $E$ band, which correspond to the quantized spin-wave
excitations on the ring.\cite{OW_SPINDYN,OW_Cr8} Observation of the spin waves in CsFe$_8$, expected at
around 7.5~meV and 10~meV at low temperatures, could thus allow further insight into this issue.

%

\section{Discussion}

In the previous sections it has been shown that the generic Hamiltonian $\hat{H}$, Eq.~(\ref{H}), which
involves only the two parameters $J$ and $D$, provides an excellent description of a significant portion of
the low-lying excitations in CsFe$_8$. In the analysis, the states were labeled by the spin quantum number
$S$, which, however, since $D$ is non-zero, is not an exact quantum number (though $M$ still is). The
question thus arises as to how good a quantum number $S$ is. The situation for CsFe$_8$ is in fact quite
ambivalent.

\begin{figure}
\includegraphics{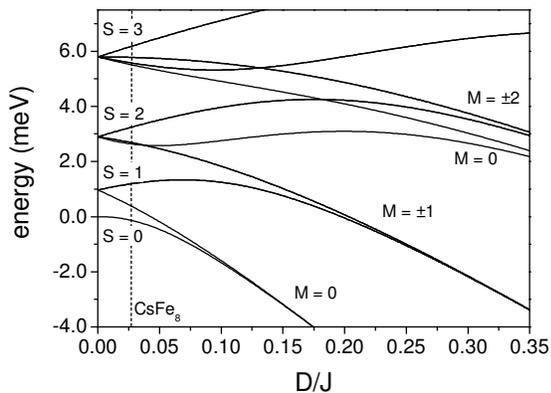}
\caption{\label{fig9} Low-lying energy spectrum of an octanuclear $s = 5/2$ wheel as function of $D/J$, as
calculated from $\hat{H}$ with $J$ = -1.80~meV. At small anisotropies, the levels can be classified by the
spin quantum numbers $S$ and $M$. For large anisotropies, $S$ loses its significance, though $M$ remains an
exact quantum number. The dashed line indicates the parameter $D/J$ as obtained for CsFe$_8$.}
\end{figure}

In our previous INS study\cite{OW_NVT_CsFe8} it has been demonstrated that the lowest energy transition
$|0,0\rangle\rightarrow|1,0\rangle$, or peak I, corresponds to the N\'eel-vector tunneling transition. Here,
the influence of the magnetic anisotropy is so strong that $S$ completely loses its significance as a quantum
number for the two involved states. The label $S$ only can be used in the sense that the states are connected
adiabatically to the corresponding states at zero magnetic anisotropy. This is visible in Fig.~\ref{fig9},
which presents the calculated energy spectrum as function of the ratio $D/J$ (using $J$ = -1.80~meV). For
small $D/J$ the states obviously can be labeled by $S$ and $M$ ("small" means here $D/J < 0.02$,
Ref.~\onlinecite{DJ}). For large $D/J$, where $S$ certainly is no longer a good quantum number, every state
nevertheless can be uniquely related back to one of the $|S,M\rangle$ states at small $D/J$. Hence $S$ can
still be used as a label, although the actual wave functions by no means are reasonably approximated by
eigenfunctions of $\hat{\textbf{S}}^2$.

The loss of the significance of $S$ with increasing $D/J$ can be inferred from the energy spectrum,
Fig.~\ref{fig9}, at least qualitatively (for a more rigorous discussion one would have to inspect the wave
functions, but here a qualitative consideration shall suffice). For $D/J$ = 0, the levels $|S,\pm M\rangle$
are degenerate since $S$ is an exact quantum number. For small values of $D/J$ the levels split, but the
states corresponding to a particular $S$ are still clustered together. For large $D/J$, however, the pattern
of the level clustering changes completely. For instance, at small $D/J$ the gap between the $S$ = 0 ground
state and the two $S$ = 1 levels is much larger than the splitting between the two $S$ = 1 levels. This
indicates that $S$ is a good quantum number and that the three states $|0,0\rangle$, $|1,0\rangle$, and
$|1,\pm1\rangle$ are approximately eigenfunctions of $\hat{S}^2$. For larger $D/J$, however, the two states
$|0,0\rangle$ and $|1,0\rangle$ come close to each other, while $|1,\pm1\rangle$ splits off and joins the
$|2,\pm1\rangle$ level. In this situation the wave functions are clearly no longer eigenfunctions of
$\hat{\textbf{S}}^2$.

It is interesting to note in Fig.~\ref{fig9} that the value of $D/J$ for this crossover apparently increases
with the energies of the spin states. Concerning the states of the $S$ = 0 and $S$ = 1 multiplets, the
crossover is at about $D/J$ = 0.02.\cite{DJ} At this value, however, the states of the next-higher lying $S$
= 2 multiplet (and the multiplets with $S >$ 2) are still clustered together; only at significantly higher
$D/J$ values are these levels split so strongly that they start to join levels coming from other spin
multiplets (the $|2,\pm1\rangle$ level approaches the $|1,\pm1\rangle$ level of the $S$ = 1 multiplet, and
$|2,0\rangle$ and $|2,\pm2\rangle$ merge with levels of the $S$ = 3 multiplet). From Fig.~\ref{fig9} it can
thus be seen that with increasing magnetic anisotropy $D/J$ the levels successively lose their $S$ character,
with larger crossover-$D/J$ values for larger $S$.

This provides a rather intuitive picture of the N\'eel-vector dynamics in AFM wheels. The situation at zero
magnetic anisotropy has been well discussed in previous
works:\cite{And52,Ber92,Lec97,OW_SPINDYN,OW_Cr8,OW_CCR} The lowest-lying excitations, which form the $L$
band, may be characterized as a quantized rotation of the N\'eel vector (that is, in a classical picture the
N\'eel vector may precess freely).

\begin{figure}
\includegraphics{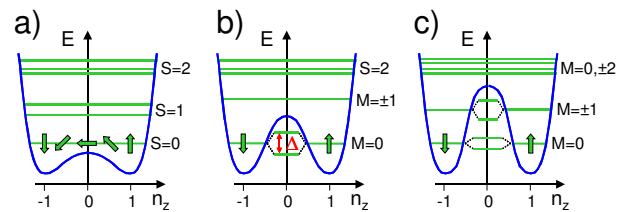}
\caption{\label{fig10} (Color online) Qualitative sketch of the semi-classical picture for the dynamics of
the N\'eel vector in AFM wheels. (a) Small magnetic anisotropy, where the dynamics is characterized by
rotation of the N\'eel vector, (b) magnetic anisotropy exceeding $S_0/\hbar > 4$, where the low-energy
dynamics is characterized by a tunneling of the N\'eel vector, and (c) larger magnetic anisotropy, where
successively further higher-lying states are driven into the tunneling regime.}
\end{figure}

The introduction of an easy-axis anisotropy means that (in a classical picture) the orientations of the
N\'eel vector along the anisotropy axis ($z$ axis) are energetically favored over the other orientations,
which is usually sketched by a double-well potential with two minima for the N\'eel vector along the $z$
axis, and an energy barrier in between (see Fig.~\ref{fig10}). For small anisotropies the energy barrier is
small and the precession of the N\'eel vector is largely undisturbed [Fig.~\ref{fig10}(a)], and thus may be
observed as for instance in the molecule Cr$_8$.\cite{OW_Cr8} With increasing anisotropy the energy barrier
increases, so that the (classical) ground state eventually falls below the top of the barrier, and the
dynamics is characterized by tunneling of the N\'eel vector, Fig.~\ref{fig10}(b).\cite{Bar90,Chi98,Mei01} In
the quantum mechanical picture, this affects the $S$ = 0 and $S$ = 1 multiplets, which lose their $S$
character. For the higher-lying multiplets, the anisotropy is not yet strong enough to compete with the
higher rotational energies involved in these states (as indicated by their larger $S$ values), so that they
are still well characterized by a rotation of the N\'eel vector, and by $S$. With further increasing
anisotropy, the two lowest-lying states are driven deeper into the tunneling regime (as indicated in
Fig.~\ref{fig9} by the exponentially shrinking gap between $|0,0\rangle$ and $|1,0\rangle$), but eventually
also the next-higher levels become tunneling states (as indicated in Fig.~\ref{fig9} by the exponentially
shrinking gap between $|1,\pm1\rangle$ and $|2,\pm1\rangle$), see Fig.~\ref{fig10}(c). This continues with
further increasing anisotropy.

Apparently for CsFe$_8$, the situation portrayed in Fig.~\ref{fig10}(b) is realized, where the low-energy
dynamics is characterized by a tunneling of the N\'eel vector (and the quantum number $S$ has lost its
meaning), while at the higher energies the dynamics can still be rationalized as a rotation of the N\'eel
vector (so that the spin multiplets with $S \geq 2$ retain their $S$ character). The $S$ = 2 multiplet
nevertheless shows an interesting signature of the strong anisotropy, namely the $|2,0\rangle$ and
$|2,\pm1\rangle$ levels are very close to each other in CsFe$_8$ (see Fig.~\ref{fig9}). This indicates a
strong fourth-order term $B^0_4$ for the $S$ = 2 multiplet with a ratio $B^0_4/D_2 = 5\times10^{-4}$ (for
comparison in Mn$_{12}$:\cite{Bir04} $B^0_4/D_2 = 5\times10^{-5}$).\cite{D2}

%

\section{Conclusions}

In this work the previous INS measurements \cite{OW_NVT_CsFe8} on the AFM wheel CsFe$_8$ were extended to a
larger temperature and energy transfer range. This has allowed us to observe the $L$-band states from the
ground state up to the $S$ = 5 multiplet. The analysis of the data, on a qualitative level, was straight
forward and of textbook quality. For the quantitative analysis the effective Hamiltonian concept, which is
well-known to be very successful in the interpretation of energy spectra, was extended such as to also
facilitate an accurate calculation of the INS scattering intensities. This allowed a least-squares fitting of
the data. The results showed that the low-energy magnetism in the CsFe$_8$ wheel is excellently described by
the generic spin Hamiltonian $\hat{H}$, Eq.~(\ref{H}), with only two magnetic parameters. The role of the
quantum number $S$ has also been discussed qualitatively, and showed that in CsFe$_8$ the situation is
ambivalent: On the one hand, the very-lowest excitations are characterized by a tunneling of the N\'eel
vector, implying that $S$ is no longer a good quantum number, while on the other hand the higher-lying states
are still well described as rotational modes of the N\'eel vector, for which $S$ is a good quantum number.

It would be clearly interesting to synthesize and identify AFM wheels with an anisotropy so strong that the
situation of Fig.~\ref{fig10}(c) is realized, i.e., where also higher-lying states are driven into a
tunneling regime. The semi-classical theory for such a situation has not been, to the best of our knowledge,
yet developed. It would be, however, important to put the qualitative picture of the N\'eel-vector dynamics
in strongly anisotropic wheels outlined in section~VI on a sound theoretical basis.

For very large $D/J$, the situation should bear a close analogy to that found in the (Mn$_4$)$_2$ molecule,
which represents a dimer of two Mn$_4$ single-molecule magnets coupled by a weak AFM
interaction.\cite{Wer02,Hil03,Sie05} This molecule attracted quite some interest recently because of its
distinct magnetic hysteresis curve. According to the effective two-sublattice Hamiltonian $\hat{H}_{AB}$,
Eq.~(\ref{Heff}), an AFM wheel can be considered magnetically as a dimer of two large spins $S_A$ and $S_B$.
Thus, for small $J$, and because of the easy-axis anisotropies of the spins $S_A$ and $S_B$, one formally
would have exactly the situation of a dimer of single-molecule magnets with a weak AFM interaction, similarly
as in (Mn$_4$)$_2$. It is thus predicted that in highly anisotropic AFM wheels phenomena such as
exchange-biased magnetic hysteresis curves could be observable.

%

\begin{acknowledgments}
Financial support by the Swiss National Science Foundation and the European Union (EC-RTN-QUEMOLNA, Contract
No. MRTN-CT-2003-504880) is gratefully acknowledged.
\end{acknowledgments}

%

\appendix
\section{}

\begin{table}[b]
\caption{\label{tab1} Results of the analysis of the $\lambda$ = 3.8~{\AA} neutron energy-loss spectra for
the temperatures 2.2~K, 9.7~K, and 17~K. Each peak in the data was least-square fitted by a Gaussian line;
the peak position, linewidth (FWHM), and intensity (in arb. units) are listed below. For further details of
the analysis see section~III and Fig.~\ref{fig5}. The last column reports the theoretically estimated
instrumental resolution of IN5 at the respective transfer energies.}
\begin{ruledtabular}
\begin{tabular}{cccccc}
 peak & $T$ & peak energy & width & intensity & instr. width \\
      &     & (meV) & (meV) & ($\times10^{-2}$~a.u.) & (meV) \\
\hline
 I  &  2.2~K & 0.5274(3)  & 0.1732(8)  & 1.31(1) & 0.132\\
    & 9.7~K & 0.5275(4)  & 0.1747(9)  & 0.761(4) & 0.132\\
    & 17~K & 0.5237(8)  & 0.186(2)   & 0.393(3) & 0.132\\
 II &  2.2~K & 1.3450(3)  & 0.1647(6)  & 1.382(5) & 0.119\\
    & 9.7~K & 1.324(3)   & 0.141(5)   & 0.5(1) & 0.119\\
    & 17~K & 1.327(4)   & 0.140(7)   & 0.24(5) & 0.119\\
 III&  2.2~K & 2.289(9)   & 0.17(2)    & 0.029(3) & 0.105\\
    & 9.7~K & 2.263(2)   & 0.210(4)   & 0.41(1) & 0.106\\
    & 17~K & 2.253(4)   & 0.221(6)   & 0.33(1) & 0.106\\
  IV& 9.7~K & 1.43(2)    & 0.20(2)    & 0.4(1) & 0.118\\
    & 17~K & 1.46(2)    & 0.22(1)    & 0.37(5) & 0.117\\
   V& 9.7~K & 2.036(4)   & 0.134(6)   & 0.09(1) & 0.109\\
    & 17~K & 2.042(3)   & 0.140(4)   & 0.13(1) & 0.109\\
  VI& 17~K & 2.937(2)   & 0.338(3)   & 0.354 & 0.097
\end{tabular}
\end{ruledtabular}
\end{table}

As described in section~III, the experimental INS spectra were analyzed by approximating the background with
an analytical function and least-square fitting of the corrected data to an appropriate number of Gaussian
lines. The resulting parameters for the six peaks I to VI visible in the 3.8~{\AA} energy-loss spectra
[Figs.~\ref{fig2}(a) and \ref{fig5}] are listed in Table~A.I, those for the peaks I, II, and IV visible in
the 5.0~{\AA} energy-loss spectra [Fig.~\ref{fig3}(a)] are listed in Table~A.II, and those for the three
peaks VI' to VIII' visible in the 3.8~{\AA} energy-gain spectra at $T$ = 60~K [Figs.~\ref{fig3}(b) and
\ref{fig6}] are listed in Table~A.III. The results of a similar analysis for the other neutron energy-gain
data, while in agreement, is not given here because of redundancy. The linewidths reported in the tables are
the widths of the best-fit Gaussian lines (no convolution with instrumental resolution). For comparison, also
the theoretically expected instrumental resolutions of IN5 are given in the last columns. These, however, are
not achieved in the experiment (probably due to sample inhomogeneity). This is evident from the experimental
resolution at the elastic line, which for $\lambda$ = 5.0~{\AA} (3.8~{\AA}) was 121~$\mu$eV (161~$\mu$eV) as
compared to the theoretical instrumental resolution of 105~$\mu$eV (141~$\mu$eV). Hence, for the neutron
energy-loss side the experimental resolution should be assumed to be essentially independent on energy
transfer and of the value at the elastic line. The errors in the tables are the statistical errors returned
by the fitting routine. For a complete error analysis instrumental uncertainties (widths of time channels,
inaccurate sample position, etc.), which amount to a few $\mu$eV, also have to be considered.

\begin{table}
\caption{\label{tab2} Results of the analysis of the $\lambda$ = 5.0~{\AA} neutron energy-loss spectra for
the temperatures 2.4~K, 9.7~K, and 17~K. Each peak in the data was least-square fitted by a Gaussian line;
the peak position, linewidth (FWHM), and intensity (in arb. units) are listed below. For further details of
the analysis see section~III. The last column reports the theoretically estimated instrumental resolution of
IN5 at the respective transfer energies.}
\begin{ruledtabular}
\begin{tabular}{cccccc}
 peak & $T$ & peak energy & width & intensity & instr. width \\
      &     & (meV) & (meV) & ($\times10^{-2}$~a.u.) & (meV) \\
\hline
I  &  2.4~K & 0.5178(5) & 0.133(1) & 2.53(2) & 0.094\\
   & 9.7~K & 0.5211(2) & 0.1255(4) & 1.361(5) & 0.094\\
   & 17~K & 0.5203(5) & 0.12(1) & 0.66(1) & 0.094\\ 
II &   2.4~K & 1.3171(4) & 0.1303(8) & 2.41(1) & 0.080\\
   &  9.7~K & 1.3040(5) & 0.110(2)  & 0.84(4) & 0.080\\
   &  17~K & 1.303(1)  & 0.101(4)  & 0.30(3) & 0.080\\ 
IV &  9.7~K & 1.398(5)  & 0.184(4)  & 0.86(4) & 0.079\\
   &  17~K & 1.420(5)  & 0.208(5)  & 0.80(3) & 0.078
\end{tabular}
\end{ruledtabular}
\end{table}

\begin{table}[t]
\caption{\label{tab3} Results of the analysis of the $\lambda$ = 5.0~{\AA} neutron energy-gain spectra at $T$
= 60~K. Each peak in the data was least-square fitted by a Gaussian line. The peak position, linewidth
(FWHM), and intensity (in arb. units) are listed below. For further details of the analysis see sections~III
and IV, and Fig.~\ref{fig6}. The last column reports the theoretically estimated instrumental resolution of
IN5 at the respective transfer energies.}
\begin{ruledtabular}
\begin{tabular}{ccccc}
 peak &  peak energy & width & intensity & instr. width \\
       & (meV) & (meV) & ($\times10^{-2}$~a.u.) & (meV)  \\
\hline
VI'   & -2.945(4) & 0.363(8) & 0.112(3) & 0.187\\
VII'  & -3.883(8) & 0.400(2) & 0.062(3) & 0.218\\
VIII' & -4.83(4)  & 0.33(9)  & 0.010(3) & 0.252
\end{tabular}
\end{ruledtabular}
\end{table}

For a given peak, the best-fit values for the transition energies obtained from the various curves show some
variation, however, this is well within the typical limits. For transition II, the obtained linewidths at the
higher temperatures, see Tables~A.I and A.II, are unrealistically small, however, this is not significant in
view of the close proximity of transitions II and IV. \\ \\

\section{}

In this appendix, explicit expressions for the symmetrized spherical tensors $U^{(k)}_q$ and the interference
factors $f^{kq}_{ij}$ are given, which were introduced via Eq.~(\ref{Ieff}) in section~IV in order to provide
a compact notation for the INS scattering intensity of powder samples. The symmetrized spherical tensors read
\begin{eqnarray}
 U^{(0)}_0({\bf \tilde{S}}_i{\bf \tilde{S}}_j) &=&
   \tilde{S}_{ix} \tilde{S}_{jx} + \tilde{S}_{iy} \tilde{S}_{jy} + \tilde{S}_{iz} \tilde{S}_{jz},
   \cr
 U^{(2)}_0({\bf \tilde{S}}_i{\bf \tilde{S}}_j) &=&
   (2 \tilde{S}_{iz} \tilde{S}_{jz} - \tilde{S}_{ix} \tilde{S}_{jx} - \tilde{S}_{iy} \tilde{S}_{jy})/\sqrt{6},
   \cr
 U^{(2)}_2({\bf \tilde{S}}_i{\bf \tilde{S}}_j) &=&
   (\tilde{S}_{ix} \tilde{S}_{jx} - \tilde{S}_{iy} \tilde{S}_{jy})/\sqrt{2},
   \cr
 U^{(2)}_{-2}({\bf \tilde{S}}_i{\bf \tilde{S}}_j) &=&
   \tilde{S}_{ix} \tilde{S}_{jy} + \tilde{S}_{iy} \tilde{S}_{jx},
   \cr
 U^{(2)}_1({\bf \tilde{S}}_i{\bf \tilde{S}}_j) &=&
   \tilde{S}_{ix} \tilde{S}_{jz} + \tilde{S}_{iz} \tilde{S}_{jx},
   \cr
 U^{(2)}_{-1}({\bf \tilde{S}}_i{\bf \tilde{S}}_j) &=&
   \tilde{S}_{iy} \tilde{S}_{jz} + \tilde{S}_{iz} \tilde{S}_{jy},
\end{eqnarray}
and the corresponding interference factors
\begin{eqnarray}
 f^{00}_{ij}(Q,{\bf R}_{ij}) &=&  F^*_i F_j {2 \over 3} j_0(Q R_{ij}),
   \cr
 f^{20}_{ij}(Q,{\bf R}_{ij}) &=&  F^*_i F_j
   j_2(Q R_{ij}){3 R^2_{ij,z} - R^2_{ij} \over  \sqrt{6} R^2_{ij}},
   \cr
 f^{22}_{ij}(Q,{\bf R}_{ij}) &=&  F^*_i F_j
   j_2(Q R_{ij}){R^2_{ij,x} - R^2_{ij,y} \over \sqrt{2} R^2_{ij}},
   \cr
 f^{2-2}_{ij}(Q,{\bf R}_{ij}) &=&  F^*_i F_j
   j_2(Q R_{ij}){R_{ij,x} R_{ij,y} \over R^2_{ij}},
   \cr
 f^{21}_{ij}(Q,{\bf R}_{ij}) &=&  F^*_i F_j
   j_2(Q R_{ij}){R_{ij,x} R_{ij,z} \over R^2_{ij}},
   \cr
 f^{2-1}_{ij}(Q,{\bf R}_{ij}) &=&  F^*_i F_j
   j_2(Q R_{ij}){R_{ij,y} R_{ij,z} \over R^2_{ij}}.
\end{eqnarray}

In the previous literature, the form factors $F_i(Q)$ were not included in the definition of the interference
factors, which is physically sound, as these two factors have different physical origins. Here, however, they
were included in the $f_{ij}^{kq}$ terms, because the approach described in section~IV for the calculation of
the INS intensity, from an effective Hamiltonian, is also very useful in the case of heteronuclear systems
(such as the single-molecule magnet Mn$_4$Br, Ref.~\onlinecite{Sie05}). In these cases the equations simplify
considerably in the chosen notation.

%

%

\begin{references}

\bibitem{Ses93}
R. Sessoli, D. Gatteschi, A. Caneschi, and M. A. Novak, Nature (London) {\bf 365}, 141 (1993).

\bibitem{Fri96}
J. R. Friedman, M. P. Sarachik, J. Tejeda, and R. Ziolo, Phys. Rev. Lett. {\bf 76}, 3830 (1996).

\bibitem{Tho96}
L. Thomas, F. Lionti, R. Ballou, D. Gatteschi, R. Sessoli, and B. Barbara, Nature (London) {\bf 383}, 145
(1996).

\bibitem{Wer99}
W. Wernsdorfer and R. Sessoli, Science {\bf 284}, 133 (1999).

\bibitem{Mn12_Fe8}
D. Gatteschi and R. Sessoli, Angew. Chem. Int. Ed. {\bf 42}, 268 (2003).

\bibitem{Taf94}
K. L. Taft, C. D. Delfs, G. C. Papaefthymiou, S. Foner, D. Gatteschi, and S. J. Lippard,
 J. Am. Chem. Soc. {\bf 116}, 823 (1994).

\bibitem{Gat94}
D. Gatteschi, A. Caneschi, L. Pardi, and R. Sessoli, Science {\bf 265}, 1054 (1994).

\bibitem{Inn05}
E. J. L. McInnes, S. Piligkos, G. A. Timco, and R. E. P. Winpenny, Coordin. Chem. Rev. \textbf{249}, 2577
(2005).

\bibitem{OW_CCR}
O. Waldmann, Coordin. Chem. Rev. \textbf{249}, 2550 (2005).

\bibitem{Cor99}
A. Cornia, M. Affronte, A. G. M. Jansen, G. L. Abbati, and D. Gatteschi, Angew. Chem. Int. Ed. Engl. {\bf
38}, 2264 (1999).

\bibitem{Cor99b}
A. Cornia, A. G. M. Jansen, and M. Affronte, Phys. Rev. B {\bf 60}, 12177 (1999).

\bibitem{OW_XFe6}
O. Waldmann, J. Sch\"ulein, R. Koch, P. M\"uller, I. Bernt, R. W. Saalfrank, H. P. Andres, H. U. G\"udel, and
P. Allenspach, Inorg. Chem. \textbf{38}, 5879 (1999).

\bibitem{OW_CsFe8}
O. Waldmann, R. Koch, S. Schromm, J. Sch\"ulein, P. M\"uller, I. Bernt, R. W. Saalfrank, F. Hampel, and E.
Balthes, Inorg. Chem. \textbf{40}, 2986 (2001).

\bibitem{Aff02}
M. Affronte, A. Cornia, A. Lascialfari, F. Borsa, D. Gatteschi, J. Hinderer, M. Horvatic, A. G. M. Jansen,
and M. H. Julien, Phys. Rev. Lett. \textbf{88}, 167201 (2002).

\bibitem{OW_Cr8}
O. Waldmann, T. Guidi, S. Carretta, C. Mondelli, and A. L. Dearden, Phys. Rev. Lett. \textbf{91}, 237202
(2003).

\bibitem{OW_NVT_CsFe8}
O. Waldmann, C. Dobe, H. Mutka, A. Furrer, and H. U. G\"udel, Phys. Rev. Lett. \textbf{95}, 057202 (2005).

\bibitem{San05}
P. Santini, S. Carretta, G. Amoretti, T. Guidi, R. Caciuffo, A. Caneschi, D. Rovai, Y. Qiu, and J. R. D.
Copley, Phys. Rev. B \textbf{71}, 184405 (2005).

\bibitem{dipdip}
Dipole-dipole interactions, which add a term $\sum_{i \neq j} \hat{\textbf{S}}_i \cdot \textbf{D}^{dip}_{ij}
\cdot \hat{\textbf{S}}_j$ to Hamiltonian~(\ref{H}), cannot be neglected in general. However, as long as one
is concerned only with the lowest lying states, as in the present work, their effects can be lumped into a
single-ion term $\sum_i \hat{S}^2_{i,z}$, that is, the anisotropy parameter $D$ should be understood in this
work as to parameterize both the dipole-dipole and ligand-field interactions. This approach is very
convenient for physical considerations where the actual origin of the anisotropy is not of much concern. For
magneto-chemical considerations, of course, such an approach would be inappropriate and the two contributions
would have to be distinguished.

\bibitem{And52}
P. W. Anderson, Phys. Rev. \textbf{86}, 694 (1952).

\bibitem{Ber92}
B. Bernu, C. Lhuillier, and L. Pierre, Phys. Rev. Lett. \textbf{69}, 2590 (1992).

\bibitem{Sch00}
J. Schnack and M. Luban, Phys. Rev B \textbf{63}, 014418 (2000).

\bibitem{OW_SPINDYN}
O. Waldmann, Phys. Rev. B \textbf{65}, 024424 (2002).

\bibitem{DJ}
For the discussion of the effect of the magnetic anisotropy the ratio $D/J$ is the relevant quantity.
However, its effect also depends on the number $N$, and spin length $s$, of the spin centers, which has to be
taken into account if one wants to compare the situation for different wheels. A better quantity is the
so-called tunneling action $S_0/\hbar = Ns \sqrt{2D/J}$, as obtained in a semi-classical
calculation.\cite{Chi98,Mei01} The semiclassical theory also gives the criterion $S_0/\hbar
> 4$ for the crossover to the N\'eel-vector tunneling regime, which for CsFe$_8$ is $D/J > 0.02$.

\bibitem{Bar90}
B. Barbara and E. M. Chudnovsky, Phys. Lett. A \textbf{145}, 205 (1990).

\bibitem{Chi98}
A. Chiolero and D. Loss, Phys. Rev. Lett. \textbf{80}, 169 (1998).

\bibitem{Mei01}
F. Meier and D. Loss, Phys. Rev. B \textbf{64}, 224411 (2001).

\bibitem{Saa97}
R. W. Saalfrank, I. Bernt, E. Uller, and F. Hampel, Angew. Chem. Int. Ed. Engl. \textbf{36}, 2482 (1997).

\bibitem{Lec97}
P. Lecheminant, B. Bernu, C. Lhuillier, L. Pierre, and P. Sindzingre, Phys. Rev. B \textbf{56}, 2521 (1997).

\bibitem{OW_FW_QT}
O. Waldmann, Europhys. Lett. \textbf{60}, 302 (2002).

\bibitem{Fur77}
A. Furrer and H. U. G\"udel, Phys. Rev. Lett. \textbf{39}, 657 (1977).

\bibitem{OW_INS}
O. Waldmann, Phys. Rev. B \textbf{68}, 174406 (2003).

\bibitem{Fur79}
A. Furrer and H. U. G\"udel, J. Magn. Magn. Mater. {\bf 14}, 256 (1979).

\bibitem{Gue85}
H. U. G\"udel, in {\it Magneto-Structural Correlations in Exchange-Coupled Systems}, edited by R. D. Willet
(Reidel, Amsterdam, 1985), p. 325.

\bibitem{OW_INS2}
O. Waldmann and H. U. G\"udel, Phys. Rev. B \textbf{72}, 094422 (2005).

\bibitem{OW_NVT_3x3}
O. Waldmann, Phys. Rev. B \textbf{71}, 094412 (2005).

\bibitem{FFILL}
P. J. Brown, in {\it Neutron Data Booklet}, edited by A. J. Dianoux and G. Lander (Institute Laue-Langevin,
Grenoble, 2001).

\bibitem{peaks}
The assignment of the various transitions forming the spectrum of Fig.~\ref{fig8} is as follows [compare also
with Fig.~\ref{fig4}(b)]. Peaks I and II correspond to the two possible transitions between the $S$ = 0 and
$S$ = 1 multiplets: $|0,0\rangle\rightarrow|1,0\rangle$ and $|0,0\rangle\rightarrow|1,\pm1\rangle$. The peaks
III, IV and V arise from the five transitions between the $S$ = 1 and $S$ = 2 multiplets. In order of
increasing energy, peak IV: $|1,\pm1\rangle\rightarrow|2,0\rangle$ and
$|1,\pm1\rangle\rightarrow|2,\pm1\rangle$, peak V: $|1,\pm1\rangle\rightarrow|2,\pm2\rangle$, peak III:
$|1,0\rangle\rightarrow|2,0\rangle$ and $|1,0\rangle\rightarrow|2,\pm1\rangle$. Finally, peak VI is produced
by six of the eight possible transitions between the $S$ = 2 and $S$ = 3 multiplets:
$|2,\pm1\rangle\rightarrow|3,0\rangle$, $|2,0\rangle\rightarrow|3,0\rangle$,
$|2,\pm1\rangle\rightarrow|3,\pm1\rangle$, $|2,0\rangle\rightarrow|3,\pm1\rangle$,
$|2,\pm2\rangle\rightarrow|3,\pm3\rangle$, and $|2,\pm1\rangle\rightarrow|3,\pm2\rangle$. Concerning the two
remaining transitions, transition $|2,\pm2\rangle\rightarrow|3,\pm1\rangle$ (at 2.30~meV) coincides with the
transition $|1,0\rangle\rightarrow|2,\pm1\rangle$ within the resolution of Fig.~\ref{fig8}(a), and transition
$|2,\pm2\rangle\rightarrow|3,\pm2\rangle$ appears as a weak satellite to peak III at 2.52~meV.

\bibitem{Cin02}
F. Cinti, M. Affronte, and A. G. M. Jansen, Eur. Phys. J. B \textbf{30}, 461 (2002).

\bibitem{Car03}
S. Carretta, J. van Slageren, T. Guidi, E. Liviotti, C. Mondelli, D. Rovai, A. Cornia, A. L. Dearden, F.
Carsughi, M. Affronte, C. D. Frost, R. E. P. Winpenny, D. Gatteschi, G. Amoretti, and R. Caciuffo, Phys. Rev.
B \textbf{67}, 094405 (2003).

\bibitem{OW_NaFe6}
O. Waldmann, R. Koch, S. Schromm, P. M\"uller, I. Bernt, and R. W. Saalfrank, Phys. Rev. Lett. \textbf{89},
246401 (2002).

\bibitem{OW_FIMEI}
O. Waldmann, C. Dobe, S. T. Ochsenbein, H. U. G\"udel, and I. Sheikin, Phys. Rev. Lett. \textbf{96}, 027206
(2006).

\bibitem{EPR}
G. Carver, O. Waldmann, C. Dobe, H. U. G\"udel, and A. L. Barra (unpublished).

\bibitem{Bir04}
R. Bircher, G. Chaboussant, A. Sieber, H. U. G\"udel, and H. Mutka, Phys. Rev. B {\bf 70}, 212413 (2004).

\bibitem{D2}
The parameters $D_2$ and $B^0_4$ relate to the effective Hamiltonian for the $S$ = 2 multiplet $\hat{H}_{S=2}
= D_2 [\hat{S}^2_z - 1/3S(S+1)] + B^0_4 \hat{O}^0_4(S)$, where $\hat{O}^0_4(S)$ is a Stevens
operator.\cite{Abr70}

\bibitem{Abr70}
A. Abragam and B. Bleaney, {\it Electron Paramagnetic Resonance of Transition Ions} (Clarendon Press, Oxford,
1970).

\bibitem{Wer02}
W. Wernsdorfer, N. Aliaga-Alcalde, D. N. Hendrickson, and G. Christou, Nature \textbf{416}, 406 (2002).

\bibitem{Hil03}
S. Hill, R. S. Edwards, N. Aliaga-Alcalde, and G. Christou, Science \textbf{302}, 1015 (2003).

\bibitem{Sie05}
A. Sieber, D. Foguet-Albiol, O. Waldmann, S. T. Ochsenbein, R. Bircher, G. Christou, F. Fernandez-Alonso, H.
Mutka, and H. U. G\"udel, Inorg. Chem. \textbf{44}, 6771 (2005).


\end{references}
\end{document}